\documentclass{cernrep}

\usepackage{varwidth}
\usepackage{xcolor}

\begin{document}
\title{AC/RF Superconductivity}
\author{G. Ciovati\footnote{gciovati@jlab.org}}
\institute{Thomas Jefferson National Accelerator Facility, Newport News, Virginia, USA}
\maketitle

\begin{abstract}
This contribution provides a brief introduction to AC/RF superconductivity, with an emphasis on application to accelerators. The topics covered include the surface impedance of normal conductors and superconductors, the residual resistance, the field dependence of the surface resistance, and the superheating field.\\

\end{abstract}

\section{Introduction}
This chapter provides an introductory-level tutorial to AC/RF superconductivity. The emphasis is on the application to resonant cavities for particle accelerators. In this respect, we will present the basic theoretical concepts and experimental results related to the low-field surface impedance, the superheating field, and the field dependence of the surface resistance. All these topics are presented to a greater depth in the bibliography and some of the references listed at the end of this tutorial.

Approximately 20 years after the discovery of superconductivity in 1911, experimental evidence of a large change in conductivity at the transition temperature was demonstrated by using Radio-Frequency (RF) currents~\cite{Silsbee1932,McLennan1932}. Shortly thereafter, a theory of the electrodynamics of superconductors, based on the phenomenological two-fluid model, was proposed by Fritz and Heinz London~\cite{London1934,London1935}. A new theory of the electrodynamics of superconductors by Mattis and Berdeen was published in 1958~\cite{Mattis1958}, based on the Bardeen--Cooper--Schrieffer (BCS) theory, which had been published one year earlier~\cite{BCS1957}. Experimental results based on far-infrared transmission through superconducting thin films and supporting the theory were published by Tinkham \textit{et al.} in the same period~\cite{Tinkham1956, Tinkham1960}.

Regarding the highest AC/RF magnetic field that can be applied to a superconductor, the so-called \textit{superheating} field, the earliest theoretical work, based on the Ginzburg--Landau (GL) theory, dates back to the 1960s~\cite{Ginzburg1958, deGennes1965, Matricon1967}. Experimental results in the range 90--300 MHz for both type I and type II superconductors in the vicinity of the critical temperature, $T_\mathrm{c}$, and consistent with the theory, were published in 1977~\cite{Yogi1977}.

Whereas niobium is the superconductor almost exclusively used to produce resonant cavities for particle accelerators, superconducting materials with higher critical temperatures are also being used for RF applications in passive microwave devices, such as filters, resonators, and antennas for mobile communications~\cite{Gallop1997}, and to produce microresonators for a variety of applications, such as photon detectors and quantum circuits~\cite{Zmuid2012}.

\section{Basics of RF cavities}
Generally speaking, a resonant cavity is any volume enclosed by metallic walls that contains oscillating electromagnetic fields.
For application to particle accelerators, the electromagnetic energy stored within the cavity is used to accelerate a charged particle beam. The frequency range relevant for accelerator applications is RF (3 kHz -- 300 GHz). 

\begin{figure}[ht]
\begin{center}
\includegraphics[width=8cm]{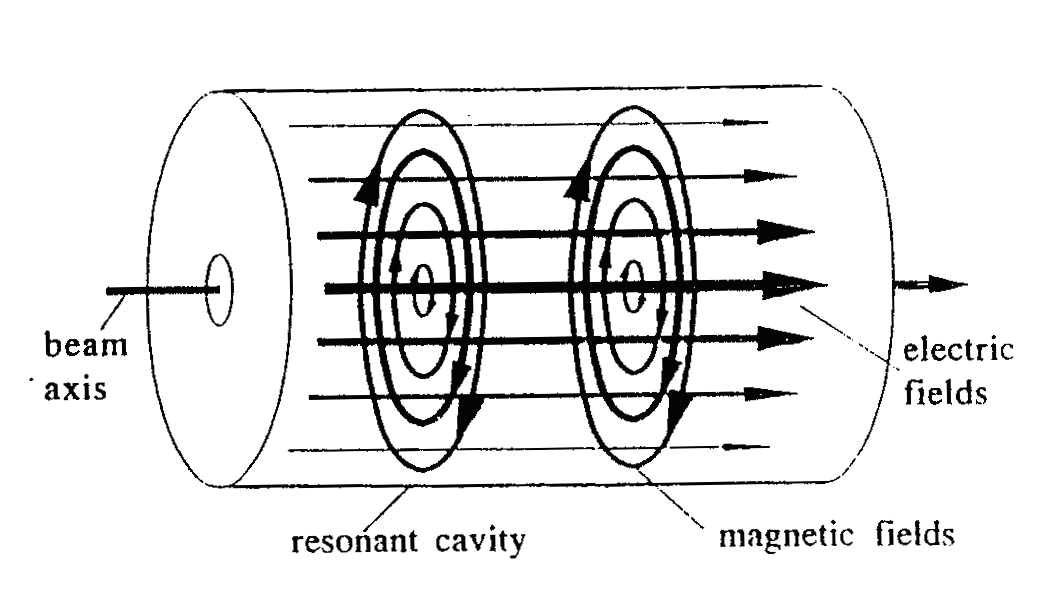}
\caption{Electric and magnetic fields for the TM$_{010}$ mode inside a pill-box cavity.}
\label{fig:cavity}
\end{center}
\end{figure}

The electromagnetic field inside an RF cavity is the solution to the wave equation:
\begin{equation}
\left(\nabla^2-\frac{1}{c^2}\frac{\partial^2}{\partial t^2}\right)\left\{\begin{array}{c}\mathbf{E}\\
       \mathbf{H}\end{array} \right\}= 0, \label{eq:a1}
\end{equation}
with the boundary conditions $\hat{n}\times\mathbf{E}=0$ and $\hat{n}\cdot\mathbf{H}=0$, where $\hat{n}$ is the unit vector normal to the surface.
Solutions to Eq.~(\ref{eq:a1}) with the specified boundary conditions can be separated into two families of resonant modes with different eigenfrequencies, based on the direction of the electric and magnetic field:
\begin{itemize}
\item TE$_{mnp}$ modes having only transverse electric fields, and
\item TM$_{mnp}$ modes having only transverse magnetic fields (but a longitudinal component of the electric field),
\end{itemize}
where $m$, $n$, and $p$ are indices denoting the number of zeros in the $\phi$, $\rho$, and $z$ directions, respectively, in cylindrical coordinates.
A useful example of a resonant cavity is a metallic cylindrical waveguide of length $L$, shorted by metallic plates at both ends. This geometry is commonly referred to as `pill-box'. The mode used to accelerate charged particles in RF cavities having a geometry resembling that of a pill-box is the TM$_{010}$. The electric and magnetic fields, as well as the resonant frequency of this mode, can be calculated analytically for the pill-box geometry:
\begin{align}
\label{eq:a2}
\begin{split}
E_z = E_0J_0\left(\frac{2.405\rho}{R}\right)\mathrm{e}^{i\omega t}, \\
H_\phi = -i \frac{E_0}{\eta}J_1\left(\frac{2.405\rho}{R}\right)\mathrm{e}^{i\omega t}, \\
\omega_0 = \frac{2.405c}{R},
\end{split}
\end{align}
where $J_0$ and $J_1$ are Bessel functions of zeroth and first order, respectively, $R$ is the pill-box radius, $c$ is the speed of light, $\omega$ is the angular frequency, and $\eta=\sqrt{\mu_0/\epsilon_0}\simeq 377$ $\Omega$ is the impedance of a vacuum.
Equation~(\ref{eq:a2}) shows that the electric field, being at a maximum on-axis, can be used to accelerate charged particles travelling along the axis of the cavity. A schematic representation of the electric and magnetic fields inside a pill-box type cavity is shown in Fig.~\ref{fig:cavity}.

Other resonant modes that are sometimes used are the TE$_{011}$ mode and the TM$_{110}$ mode. The first of these has a zero electric field on the cavity surface and is used to study the surface resistance of superconductors in RF magnetic fields. The second has a transverse component of the electric field on axis, tilting the beam, which is sometimes necessary in collider accelerators in order to provide a head-on collision between two beams and thereby increase the luminosity. The deflecting TM$_{110}$ mode has also been used in an SRF separator cavity to separate beams of different particles~\cite{Citron1979}.

Although the resonant frequency of the TM$_{010}$ mode does not depend on the pill-box length, \textit{L}, the following condition for synchronism between the beam and the electric field in the cavity sets the cavity length:
\begin{equation}
L=\beta c\frac{T_\mathrm{RF}}{2},  \label{eq:a3}
\end{equation}
where $\beta$ is the speed of the particle relative to the speed of light and $T_\mathrm{RF}=2\pi/\omega_0$ is the period of oscillation of the RF field.
The condition imposed by Eq.~(\ref{eq:a3}) assures that, as an example, a bunch of relativistic electrons entering the cavity at time $t = 0$, when $E_z=0$, will experience the maximum acceleration as they travel along the cavity axis.

\subsection{Figures of merit}

The accelerating field of the cavity, $E_\mathrm{acc}$, is defined as the ratio of the accelerating voltage, $V_\mathrm{c}$, divided by the cavity length. $V_\mathrm{c}$ is obtained by integrating the electric field at the particle's position as it traverses the cavity:
\begin{equation}
E_\mathrm{acc}=\frac{V_\mathrm{c}}{L} =\frac{1}{L}\left|\int_0^L E_z(\rho=0,z)\mathrm{e}^{i\omega_0z/c} \,\mathrm{d}z\right|.  \label{eq:a4}
\end{equation}
Other important parameters are the ratios of the peak electric and magnetic fields on the cavity surface divided by the accelerating field, $E_\mathrm{p}/E_\mathrm{acc}$ and $B_\mathrm{p}/E_\mathrm{acc}$, respectively, as they are related to practical limitations of a cavity's performance, such as field emission and quench.

\begin{figure}[ht]
\begin{center}
\includegraphics[width=8cm]{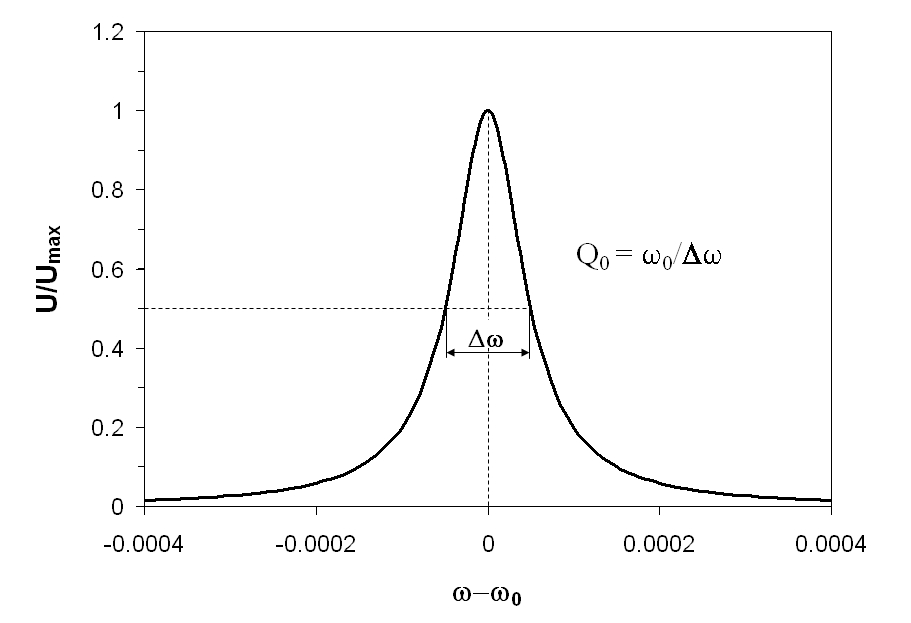}
\caption{The normalized stored energy as a function of frequency for a resonator with resonant frequency $\omega_0$ and $Q_0=10^4$.}
\label{fig:resonance}
\end{center}
\end{figure}

The power dissipated as heat in the cavity wall, $P_\mathrm{c}$, and the energy stored within its volume, \textit{U}, are given by 
\begin{equation}
P_\mathrm{c}=\frac{1}{2} \Re \left\{ \int_V \mathbf{J}\cdot\mathbf{E} \,\mathrm{d}v \right\} = \frac{1}{2} \int_S R_\mathrm{s}|\mathbf{H}|^2 \,\mathrm{d}a, \label{eq:a5}
\end{equation}
\begin{equation}
U=\frac{1}{2} \mu_0 \int_V |\mathbf{H}|^2 \,\mathrm{d}v.  \label{eq:a6}
\end{equation}
The quality factor of the cavity, $Q_0$, is defined, in the same way as for any resonator, as the ratio of the energy stored divided by the energy dissipated in in one RF period:
\begin{equation}
Q_0=\frac{\omega_0 U}{P_\mathrm{c}} =\frac{\omega_0 \mu_0 \int_V |\mathbf{H}|^2 \,\mathrm{d}v}{\int_S R_\mathrm{s}|\mathbf{H}|^2 \,\mathrm{d}a}.  \label{eq:a7}
\end{equation}
$Q_0$ can be calculated from the Breit--Wigner resonance curve as the ratio of the resonant frequency, divided by the full width at half maximum, as shown in Fig.~\ref{fig:resonance}.

Assuming that the surface resistance is uniform over the cavity surface and does not depend on the amplitude of the applied field, it is possible to define from Eq.~(\ref{eq:a7}) a geometry factor, \textit{G}, that depends only on the cavity shape (but not its size) and that provides a direct relation between $Q_0$ and $R_\mathrm{s}$:
\begin{equation}
G =\frac{\omega_0 \mu_0 \int_V |\mathbf{H}|^2 \,\mathrm{d}v}{\int_S |\mathbf{H}|^2 \,\mathrm{d}a} = Q_0R_\mathrm{s}.  \label{eq:a8}
\end{equation}
The assumptions on the definition of \textit{G} are usually valid at low field amplitudes.

The figures of merit for the TM$_{010}$ mode in a pill-box cavity, calculated from the analytical fields of Eq.~(\ref{eq:a2}), are as follows:
\begin{align}
\label{eq:a9}
\begin{split}
& E_\mathrm{acc} = \frac{2}{\pi} E_0, \\
& E_\mathrm{p}/E_\mathrm{acc} = \frac{\pi}{2} = 1.57,\\
& H_\mathrm{p}/E_\mathrm{acc} = \frac{\pi}{2} \frac{J_1(1.84)}{\eta} = 2430 \frac{\mathrm{A}\cdot\mathrm{m}^{-1}}{\mathrm{MV}\cdot\mathrm{m}^{-1}} = 30.5 \frac{\text{Oe}}{\mathrm{MV}\cdot\mathrm{m}^{-1}}, \\
& G = \eta \frac{2.405L}{2(R+L)} = \frac{453L/R}{1+L/R} \Omega .
\end{split}
\end{align}
Practical RF cavities have more complex shapes than a simple pill-box, and require numerical solvers to calculate the electric and magnetic fields inside the cavity.

\subsection{SRF cavity performance}
\begin{figure}[ht]
\begin{center}
\includegraphics[width=8cm]{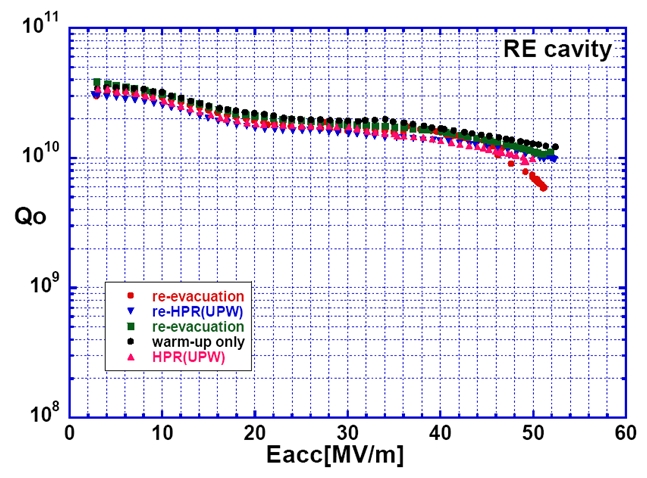}
\caption{The state-of-the-art performance of a bulk Nb cavity at 1.3~GHz and 2.0~K~\cite{Furuta2006}.}
\label{fig:performance}
\end{center}
\end{figure}

The performance of a superconducting RF cavity is described by a plot of the quality factor as a function of the accelerating gradient. The state-of-the-art performance of a 1.3 GHz bulk Nb cavity tested multiple times at 2.0 K is shown in Fig.~\ref{fig:performance} \cite{Furuta2006}. The $Q_0$ value at low field corresponds to a surface resistance of $\simeq~8~\text{n}\Omega$, and the maximum $E_\mathrm{acc}$ value corresponds to a peak surface magnetic field $B_\mathrm{p}=\mu_0H_\mathrm{p}~\simeq 185$~mT. In the following sections, we will provide a basic description of what determines the $R_\mathrm{s}$ at low field as well as the maximum obtainable $B_\mathrm{p}$ value.

\section{Surface impedance}
The electromagnetic response of a metal, whether normal or superconducting, is described by a complex surface impedance defined as follows:
\begin{equation}
Z_\mathrm{s}=\frac{|E_\parallel|}{\int_0^{\infty} J(x) \,\mathrm{d}x} = \frac{|E_\parallel|}{|H_\parallel|} = R_\mathrm{s} + iX_\mathrm{s},  \label{eq:a10}
\end{equation}
where $R_\mathrm{s}$ is the surface resistance and $X_\mathrm{s}$ is the surface reactance. In Eq.~(\ref{eq:a10}), we have neglected the displacement current, which is reasonable for good conductors at frequencies less than $\sim10^{16}$~Hz.

\subsection{The electrodynamics of normal conductors}
The electrodynamics of normal conductors is based on Maxwell's equations and on material equations that specify the relations: between the electric displacement, \textit{D}, and the electric field; between the induction field, \textit{B}, and the magnetic field; and between the current density and the electric field. Since the inside volume of a cavity is typically under vacuum, we have $D=\epsilon_0E$, $B=\mu_0H$. A relation between \textit{J} and \textit{E} can be obtained from Drude's model of `nearly free electrons' in a solid and resulting in Ohm's law:
\begin{equation}
J=\frac{ne^2}{m\tau}\frac{1}{1+i\omega \tau} E = \sigma E,  \label{eq:a11}
\end{equation}
where $\sigma=ne^2/m\tau$ is the conductivity and $\tau\simeq 10^{-14}$~s is the scattering time of the electrons, given by the ratio of the mean free path, $\ell$, divided by the Fermi velocity, $v_\mathrm{F}$. In Eq.~(\ref{eq:a11}), we have used the approximation $\omega\tau \ll 1$, which is valid at RF frequencies.
Using Maxwell's equations with the material equations, the following equation for the magnetic field inside a metal is obtained:
\begin{equation}
\nabla^2H=i\sigma\mu_0\omega H.
\label{eq:a12}
\end{equation}
The solution of Eq.~(\ref{eq:a12}) for a semi-infinite slab occupying the positive-\textit{x} region of space with a magnetic field of amplitude $H_0$ applied in the \textit{y}-direction is given by
\begin{equation}
H_y(x,t)=H_0\mathrm{e}^{-x/\delta}\mathrm{e}^{-i(x/\delta-\omega t)},
\label{eq:a13}
\end{equation}
which describes a wave propagating in the positive-\textit{x} direction with an exponentially decreasing amplitude. $\delta$ is the characteristic decay length, called the `skin depth', and given by
\begin{equation}
\delta=\sqrt{\frac{2}{\mu_0\sigma\omega}}.
\label{eq:a14}
\end{equation}
The electric field, obtained from Ampere's law, has a behaviour similar to that of the magnetic field:
\begin{equation}
E_z(x,t)=-\frac{1+i}{\sigma\delta} H_y(x,t).
\label{eq:a15}
\end{equation}
The surface impedance is then given by
\begin{equation}
Z_\mathrm{s}=\frac{|E_z(x=0)|}{H_y(x=0)} = \frac{1+i}{\sigma\delta}.
\label{eq:a16}
\end{equation}
Therefore, $R_\mathrm{s}=X_\mathrm{s}=1/\sigma\delta$.
If we consider, for example, copper ($\sigma \simeq 5.8\times 10^7~\mathrm{S}\cdot\mathrm{m}^{-1}$) at 1.5~GHz and 300~K, we obtain $\delta = 1.7~\mu$m and $R_\mathrm{s} = 10$~m$\Omega$.

At lower temperatures, the conductivity increases and therefore $\delta$ decreases and may become shorter than the electrons' mean free path. This implies that the local relation between field and current given by Ohm's law is no longer valid at low temperatures, since the distance over which the field varies becomes less than the mean free path.
A new relationship was introduced by Reuter and Sondheimer in 1948~\cite{Reuter1948}, in which \textit{J} is related to \textit{E} over a volume of the size of the mean free path:
\begin{equation}
\mathbf{J}(\mathbf{r},t)=\frac{3\sigma}{4\pi \ell}\int_V \frac{\mathbf{R}[\mathbf{R}\cdot \mathbf{E}(\mathbf{r'},t-R/v_F)]}{R^4} \mathrm{e}^{-R/\ell} \,\mathrm{d}\mathbf{r'},
\label{eq:a17}
\end{equation}
with $\mathbf{R} = \mathbf{r'} - \mathbf{r}$.
An effective conductivity, $\sigma_\mathrm{eff} \approx (\delta/\ell)\sigma$, results from Eq.~(\ref{eq:a17}), showing that, unlike the DC case, increasing the purity of the metal does not improve its conductivity. This phenomenon is commonly referred to as the `anomalous skin effect'.
The surface resistance in the extreme anomalous limit ($\ell \to \infty$), valid for very good conductors such as copper at low temperatures, is given by
\begin{equation} 
R_\mathrm{n}(\ell \to \infty)=\left[ \sqrt{3} \left( \frac{\mu_0}{4\pi} \right)^2 \right]^{1/3}\omega^{2/3}(\rho \ell)^{1/3}.
\label{eq:a18}
\end{equation}
The product $\rho \ell$ is a material constant and it is $6.8\times 10^{-16}~\Omega\cdot\mathrm{m}^{-2}$ for copper. If one were to operate a 1.5 GHz copper cavity at cryogenic temperatures such as, for example, 4.2~K, rather than 300~K, the surface resistance would decrease by a factor of $\approx0.14$, which is not sufficient to justify the cost of a refrigerator.

\subsection{The electrodynamics of superconductors}
\subsubsection{Theory}
Unlike the DC case, superconductors in RF fields do not have zero resistance at finite temperatures. This is because a time-dependent magnetic field within the penetration depth generates an electric field (Faraday's law) that acts on normal electrons, as they are not shielded from it by the superconducting electrons (which form `Cooper pairs' of mass twice that of a single electron) due to their inertia.

A simple way to describe the electrodynamics of superconductors was given by the London brothers in 1934, based on the phenomenological `two-fluid' model of Gorter and Casimir~\cite{Gorter1934}. According to the model, the charge carriers are divided into two subsystems: superconducting carriers of density $n_\mathrm{s}$ and normal electrons of density $n_\mathrm{n}$. The superconducting carriers were associated later on with Cooper pairs of charge $-2e$ and mass $2m_\mathrm{e}$~\cite{Cooper1956}. The normal current $J_\mathrm{n}$ and the supercurrent $J_\mathrm{s}$ are assumed to flow in parallel, and the total current is the sum of $J_\mathrm{s}$ and $J_\mathrm{n}$. $J_\mathrm{s}$ flows with no resistance and follows the London equations:
\begin{equation}
\frac{\partial}{\partial t} \mathbf{J}_\mathrm{s} = \frac{1}{\mu_0 \lambda_\mathrm{L}^2} \mathbf{E},
\label{eq:a19}
\end{equation}
\begin{equation}
\nabla \times \mathbf{J}_\mathrm{s} = - \frac{1}{\lambda_\mathrm{L}^2} \mathbf{H},
\label{eq:a20}
\end{equation}
where $\lambda_\mathrm{L}=\sqrt{m/\mu_0n_\mathrm{s}e^2}$ is the so-called London penetration depth. Equation~(\ref{eq:a19}) (the first London equation) implies perfect conductivity, since a current would flow indefinitely in a superconductor even for zero electric field, and that an electric field is required to maintain an RF current. Equation~(\ref{eq:a20}) (the second London equation) implies the spontaneous flux exclusion from the bulk of a superconductor and that an induction field is the source of the supercurrent.
Note that Eq.~(\ref{eq:a20}) can be written as follows:
\begin{equation}
\mathbf{J}_\mathrm{s} = - \frac{1}{\lambda_\mathrm{L}^2}\mathbf{A},
\label{eq:a21}
\end{equation}
where \textbf{A} is the magnetic vector potential. Equation~(\ref{eq:a21}) represents a local condition between current and field that is valid if $\xi_0 \ll \lambda_\mathrm{L}$ or $\ell \ll \lambda_\mathrm{L}$. $\xi_0$ is the coherence length, representing the distance between electrons forming a Cooper pair.

In the presence of an RF current, $J=J_0\mathrm{e}^{i\omega t}$, the relation between \textit{J} and \textit{E} in a superconductor, is given by
\begin{equation}
J = J_\mathrm{n} + J_\mathrm{s} = (\sigma_1 - i\sigma_2)E,
\label{eq:a22}
\end{equation}
where $\sigma_1=n_ne^2/m\tau$ is the conductivity of the normal component (the same as the normal-state conductivity) and $\sigma_2= n_se^2/m\omega$ is the conductivity of the superconducting component, obtained using Eq.~(\ref{eq:a20}). The electrodynamics of the superconductor become analogous to those of a normal conductor if one replaces $\sigma$ with $\sigma_1-i\sigma_2$ in the expressions for the skin depth, Eq.~(\ref{eq:a14}), and the magnetic field, Eq.~(\ref{eq:a13}).  The skin depth becomes
\begin{equation}
\delta = \sqrt{\frac{2}{\mu_0\omega(\sigma_1-i\sigma_2)}} \simeq (1+i)\lambda_\mathrm{L} \left( 1+i\frac{\sigma_1}{2\sigma_2} \right),
\label{eq:a23}
\end{equation}
where we have made use of the approximation $\sigma_1 \ll \sigma_2$, which is valid for superconductors at temperature $T \ll T_\mathrm{c}$.
The magnetic field inside the superconductor becomes
\begin{equation}
H_y(x,t)=H_0\mathrm{e}^{-x/\lambda_\mathrm{L}}\mathrm{e}^{-i \left( \frac{x}{\lambda_\mathrm{L}} \frac{\sigma_1}{2\sigma_2} -\omega t \right) },
\label{eq:a24}
\end{equation}
and the amplitude of the magnetic field decreases exponentially with a characteristic length $\lambda_\mathrm{L}$. For niobium, $\lambda_\mathrm{L} \simeq 36$~nm, much shorter than the skin depth for copper at 1.5~GHz.
The surface impedance of a superconductor can be obtained by substituting Eq.~(\ref{eq:a23}) into Eq.~(\ref{eq:a16}). After some calculus involving complex numbers and with the approximation $\sigma_1 \ll \sigma_2$, we obtain the following:
\begin{equation}
Z_\mathrm{s}=\frac{1}{2} \mu_0^2 \omega^2 \sigma_1 \lambda_\mathrm{L}^3 + i \omega \mu_0 \lambda_\mathrm{L} .
\label{eq:a25}
\end{equation}
The equivalent circuit for a superconductor is a resistor of resistance $R_\mathrm{s}$ in parallel with an inductor of inductance $L_\mathrm{s}~=~\mu_0 \lambda_\mathrm{L}$, the so-called `kinetic inductance', due to the superconducting charge carriers.
The dependence of $R_\mathrm{s}$ on $\omega$ and $\ell$ in Eq. (\ref{eq:a25}) indicates the following.
\begin{itemize}
\item The surface resistance increases quadratically with frequency, and therefore low-frequency cavities should be considered to reduce the dissipated power.
\item The surface resistance increases with increasing purity of the material. An intuitive way to think about this is that normal-conducting electrons of higher conductivity draw a relatively higher fraction of the total current.
\end{itemize}
To understand the temperature dependence of the surface resistance, we consider the temperature dependence of $n_\mathrm{s}$ and $n_\mathrm{n}$:
\begin{equation}
\lambda_\mathrm{L}(T)^2 \propto 1/n_\mathrm{s}(T) \propto 1/[1-(T/T_\mathrm{c})^4],
\label{eq:a26}
\end{equation}
\begin{equation}
\sigma_1(T) \propto n_\mathrm{n}(T) \propto \mathrm{e}^{-\Delta/k_\mathrm{B} T}.
\label{eq:a27}
\end{equation}
Equation~(\ref{eq:a27}), which is valid for $T \ll T_\mathrm{c}$, reflects the creation of normal electrons due to the thermal break-up of Cooper pairs. From Eqs.~(\ref{eq:a27}) and (\ref{eq:a25}), we obtain the following dependence of $R_\mathrm{s}$ on frequency, material purity, and temperature:
\begin{equation}
R_\mathrm{s} \propto \omega^2 \lambda_\mathrm{L}^3 \ell \mathrm{e}^{-\Delta/k_\mathrm{B} T}, \mbox{\hspace{10 mm} for $T < T_\mathrm{c}/2$.}
\label{eq:a28}
\end{equation}
The exponential decrease of $R_\mathrm{s}$ with temperature indicates that low-temperature operation, such as 2.0~K for Nb cavities, is preferable to reduce RF losses.

Similarly to the anomalous skin effect in normal conductors, if $\xi_0 \gg \lambda_\mathrm{L}$ and $\ell \gg \lambda_\mathrm{L}$, the local relation between current and field is no longer valid. In 1953, Pippard~\cite{Pippard1953} proposed replacing Eq.~(\ref{eq:a21}) with Eq.~(\ref{eq:a29}) below, $\xi_0$ playing a role analogous to that of $\ell$ in the non-local electrodynamics of normal conductors:
\begin{equation}
\mathbf{J}(\mathbf{r})=\frac{3}{4\pi \xi_0 \lambda_\mathrm{L}^2}\int_V \frac{\mathbf{R}[\mathbf{R}\cdot \mathbf{A}(\mathbf{r'})]}{R^4} \mathrm{e}^{-R/\xi} \,\mathrm{d}\mathbf{r'},
\label{eq:a29}
\end{equation}
with $\mathbf{R} = \mathbf{r'} - \mathbf{r}$ and $1/\xi \simeq 1/\xi_0 + 1/\ell$.
The dependence of the penetration depth on the mean free path can be approximated as $\lambda \approx \lambda_\mathrm{L} \sqrt{1+\xi_0/\ell}$ and the dependence of $R_\mathrm{s}$ on material purity becomes
\begin{equation}
R_\mathrm{s} \propto \left( 1 + \frac{\xi_0}{\ell} \right)^{3/2} \ell .
\label{eq:a30}
\end{equation}
Equation~(\ref{eq:a30}) shows that $R_\mathrm{s} \propto \ell$ increases with increasing mean free path if $\ell \gg \xi_0$ (the so-called `clean limit') as discussed above, and that $R_\mathrm{s} \propto \ell^{-1/2}$ increases with decreasing mean free path if $\ell \ll \xi_0$ (the so-called `dirty limit'). Therefore $R_\mathrm{s}(\ell)$ has a minimum at $\ell = \xi_0/2$.
In 1958, Mattis and Bardeen~\cite{Mattis1958} obtained from the BCS theory a non-local equation between the total current density (including the supercurrent and the normal current) and the vector potential:
\begin{equation}
\mathbf{J}(\mathbf{r})=\frac{3}{4\pi^2 v_\mathrm{F} \hbar \lambda_\mathrm{L}^2}\int_V \frac{\mathbf{R}[\mathbf{R}\cdot \mathbf{A}(\mathbf{r'}) I(\omega, R, T) \mathrm{e}^{-R/\ell}]}{R^4} \,\mathrm{d}\mathbf{r'},
\label{eq:a31}
\end{equation}
where $I(\omega, R, T)$ is a function that decays over a characteristic length $R \sim \xi_0$. Equation~(\ref{eq:a31}) becomes a product of a kernel function $K(q)$ times the vector potential in the Fourier domain: $J(q)=-K(q)A(q)$ in one dimension. The surface impedance can be obtained as a function of the kernel $K(q)$, as follows:
\begin{equation}
Z_\mathrm{s} = \frac{i \mu_0 \omega \pi}{\int_0^\infty \ln (1+\frac{K(q)}{q^2}) \,\mathrm{d}q},
\label{eq:a32}
\end{equation}
for diffuse scattering of electrons at the metal surface, such as for the case of a `rough' surface on a scale of the mean free path, or
\begin{equation}
Z_\mathrm{s} = \frac{i \mu_0 \omega}{\pi} \int_{-\infty}^\infty \frac{1}{q^2+K(q)} \,\mathrm{d}q,
\label{eq:a33}
\end{equation}
for specular reflection of electrons at the metal surface.
The calculation of the real and imaginary parts of $K(q)$ involves the solution of complex integrals, which can only be done numerically. A computer code that allows the BCS surface impedance to be calculated using Eqs.~(\ref{eq:a32}) and (\ref{eq:a33}) was written by Halbritter in 1970 \cite{Halb1970} and a copy is available from the author. An online calculator is also available~\cite{Liepe}. An analytical approximation of the BCS surface resistance valid in the local limit, for $T<T_c/2$ and $\omega~<~\Delta/\hbar$, is given by \cite{Gurevich2012}
\begin{equation}
R_\mathrm{s} \simeq \frac{\mu_0^2 \omega^2 \lambda^3 \sigma_1 \Delta}{k_\mathrm{B} T} \ln \left(\frac{2.246k_\mathrm{B} T}{\hbar \omega} \right) \exp \left(-\frac{\Delta}{k_\mathrm{B} T} \right).
\label{eq:a34}
\end{equation}
Considering the case of niobium ($\lambda = 40$~nm, $\sigma_1 = 3.3\times 10^8~\mathrm{S}\cdot\mathrm{m}^{-1}$, $\Delta/k_\mathrm{B} T_\mathrm{c} = 1.85$, $T_\mathrm{c} = 9.25$~K) at 2.0 K and 1.5 GHz, we obtain $R_\mathrm{BCS} \simeq 20$~n$\Omega$ and $X_\mathrm{s} \simeq 0.47$~m$\Omega$. The ratio of $R_\mathrm{BCS}$ for Nb at 2.0~K divided by $R_\mathrm{s}$ for Cu at 300~K is $\simeq 2 \times 10^{-6}$. Even when the Carnot efficiency $\eta_\mathrm{C}~=~0.67\%$, due 2.0~K operation and the technical efficiency of a cryoplant, $\eta_\mathrm{T}~\simeq~20\%$, are included, the reduction in power consumption by using superconducting cavities instead of normal-conducting ones is still quite significant ($\simeq 10^3$ reduction factor).

\subsubsection{Experimental results}
\begin{figure}[ht]
\begin{center}
\includegraphics[width=5cm]{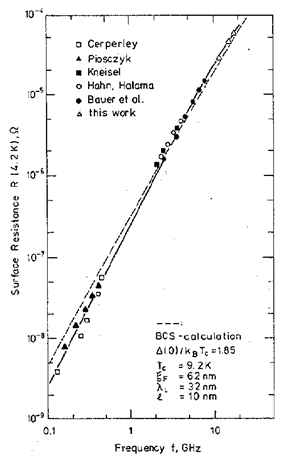}
\caption{The surface resistance measured in bulk Nb at 4.2~K as a function of frequency. The solid line is a fit to the experimental data, while the dashed line results from a calculation based on the BCS theory. Reprinted from~\cite{Halb1983} with permission from IEEE.}
\label{fig:freq_dep}
\end{center}
\end{figure}

Figure \ref{fig:freq_dep} shows the surface resistance measured in bulk Nb at 4.2 K over a broad range of frequencies~\cite{Halb1983}. Small deviations from the BCS theory can be explained by strong coupling effects or an anisotropic energy gap, in the presence of scattering by impurities or inhomogeneities.

\subsubsection{Residual resistance}
\begin{figure}[ht]
\begin{center}
\includegraphics[width=8cm]{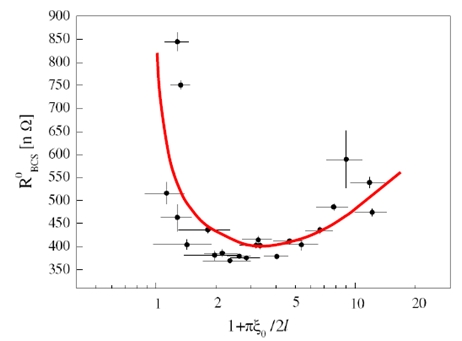}
\caption{The surface resistance measured in Nb thin films at 4.2~K and 1.5~GHz as a function of a parameter related to the mean free path. The solid line is a result from a calculation based on the BCS theory. Reprinted from~\cite{Benvenuti1999} with permission from Elsevier.}
\label{fig:mat_dep}
\end{center}
\end{figure}

\begin{figure}[ht]
\begin{center}
\includegraphics[width=8.6cm]{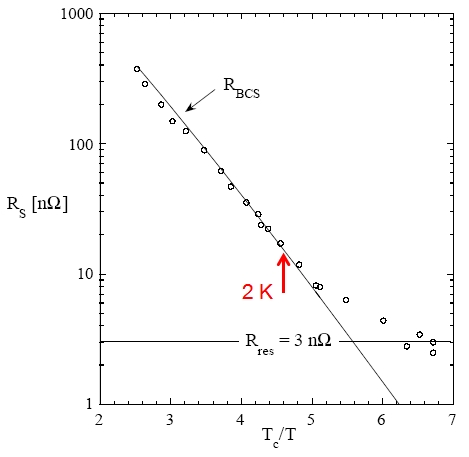}
\caption{The temperature dependence of the low-field surface resistance of bulk Nb at 1.3~GHz, showing a saturation to a $T$-independent value at low temperatures~\cite{Aune2000}.}
\label{fig:T_dep}
\end{center}
\end{figure}

\begin{figure}[ht]
\begin{center}
\includegraphics[width=8.7cm]{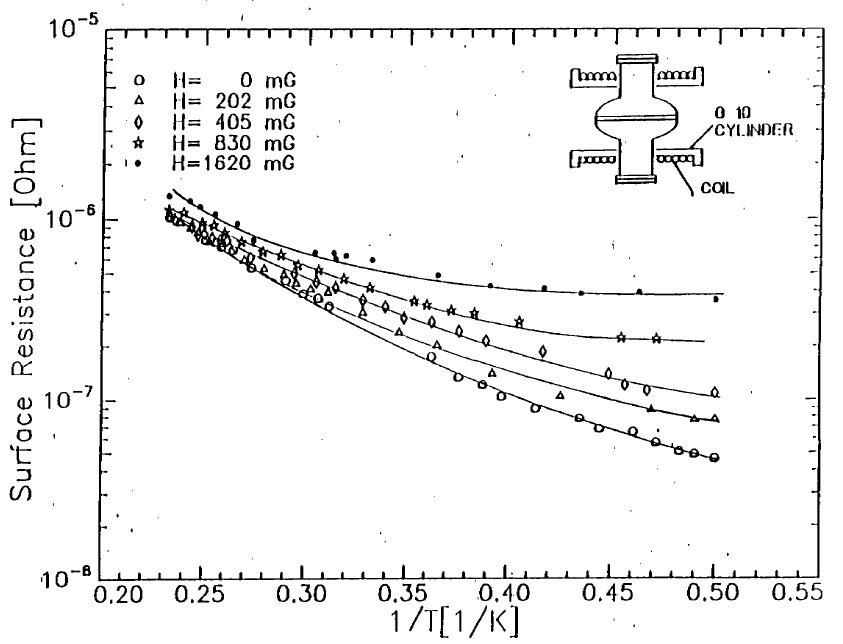}
\caption{The low-field surface resistance as a function of temperature for bulk Nb at 1.5~GHz for different amplitudes of a DC magnetic field applied during the cavity cool-down. The solid lines are fitted with $R_\mathrm{s}(T)=R_\mathrm{BCS}(T)+R_\mathrm{res}$~\cite{Kneisel1994}.}
\label{fig:Rres_H}
\end{center}
\end{figure}

\begin{figure}[ht]
\begin{center}
\includegraphics[width=14cm]{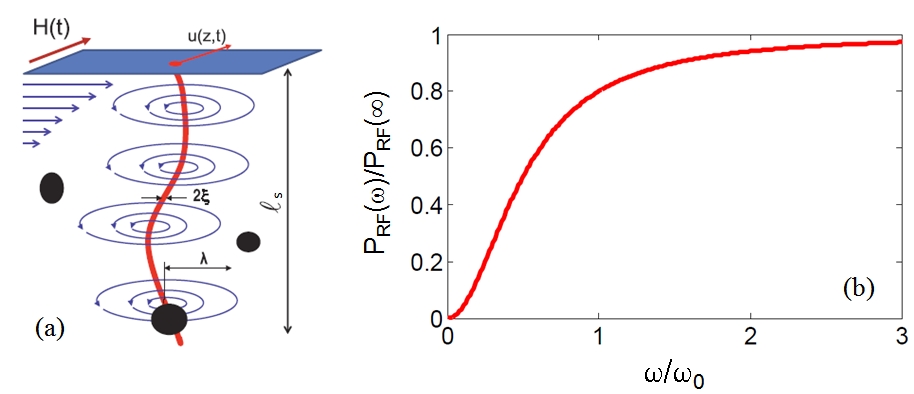}
\caption{(a) A schematic representation of a pinned fluxoid near the surface of a superconductor in the presence of an RF field. (b) The normalized power dissipation due to the oscillatory motion of a fluxoid, as a function of frequency normalized to $\omega_0=\epsilon_0/2\eta\lambda^2$, where $\epsilon_0$ is the fluxoid line energy and $\eta$ is the viscous drag coefficient~\cite{Gurevich2013}.} 
\label{fig:fluxoid}
\end{center}
\end{figure}

Improvements in the surface preparation of bulk Nb cavities over the past 40 years have reduced the typical residual resistance value from $\sim 100$~n$\Omega$ to $\sim 1-10$~n$\Omega$. $R_\mathrm{res}$ becomes the dominant term in the surface resistance at low frequency ($<\sim~750$~MHz) and low temperatures ($<\sim~2.1$~K), where $R_\mathrm{BCS}$ becomes exponentially small.
There are several possibilities contributing to the residual resistance. Among those there are:
\begin{itemize}
\item losses due to trapped magnetic field,
\item losses due to normal-conducting precipitates near the surface,
\item grain boundary losses,
\item metal/oxide interface losses, and
\item losses due to normal-conducting electrons in subgap states.
\end{itemize}
Figure \ref{fig:mat_dep} shows the surface resistance of a Nb thin film deposited on Cu,  measured at 4.2~K and 1.5~GHz, as a function of a parameter related to the purity of the film \cite{Benvenuti1999}. The data are consistent with the existence of a broad minimum of $R_\mathrm{BCS}(\ell)$, as predicted by calculations based on the BCS theory.
A measurement of the temperature dependence of the low-field surface resistance of bulk Nb at 1.3~GHz is shown in Fig.~\ref{fig:T_dep} \cite{Aune2000}. The data show a clear deviation from the exponential dependence as the temperature decreases towards 0~K. This additive, $T$-independent contribution to the surface resistance is the so-called residual resistance, $R_\mathrm{res}$ (not to be confused with the DC residual resistivity $\rho_\mathrm{n}=1/\sigma_1$). 
A well-known contribution to $R_\mathrm{res}$ is due to trapped DC magnetic field, due to the incomplete Meissner effect in technical materials. Figure~\ref{fig:Rres_H} shows, for example, the low-field $R_\mathrm{s}(T)$ measured for different DC magnetic field amplitudes applied while cooling the cavity down from 300~K to 4.2~K \cite{Kneisel1994}.
Any DC magnetic field at the cavity location can be trapped in the form of fluxoids pinned by defects in the material, as the cavity is cooled below $T_\mathrm{c}$. A simple estimate of this contribution can be made by assuming that all of the DC field, $H_\mathrm{ext}$, is trapped in the form of $N$ fluxoids, each carrying one flux quantum $\phi_0$, and that their normal-conducting cores, each of radius $\sim \xi_0$, dissipate according to the normal state surface resistance. An improved description was given by Gurevich, in which both the pinning strength and the dissipation due to the motion of the fluxoid's core under the RF field are taken into account \cite{Gurevich2013}. Figure~\ref{fig:fluxoid} shows a schematic representation of a pinned fluxoid with a segment of length $\ell_\mathrm{s}$ almost normal to the surface of the superconductor. At low RF frequency, the surface RF Meissner current causes rocking of the whole fluxoid segment. As the RF frequency increases, the RF oscillations of the segment are mostly localized closer and closer towards the surface, until only a tip of length $\sim\lambda$ vibrates in the high-frequency limit. The RF penetration depth and the pinned fluxoid segment set the frequency scales. The RF power dissipation due to the fluxoid's motion is given by
\begin{equation}
P_\mathrm{RF} = \left\langle \int_0^{\ell_\mathrm{s}} \dot{u}(z, t)F(z,t) \,\mathrm{d}z \right\rangle_\omega = - \frac{\omega F}{2} \mbox{Im}\int_0^{\ell_\mathrm{s}} u(z, \omega)\mathrm{e}^{-z/\lambda} \,\mathrm{d}z ,
\label{eq:a35}
\end{equation}
where $u(z, t)$ is the fluxoid displacement, obtained by solving the equation of motion, and $F=\phi_0 H_0/\lambda$ is the amplitude of the driving force from the RF field. The result from a calculation of $P_\mathrm{RF}(\omega)$ with Eq.~(\ref{eq:a35}) is shown in Fig.~\ref{fig:fluxoid}, indicating a change from $\sim \omega^2$ dependence at low frequency to $\sim \sqrt{\omega}$ at intermediate frequency and frequency independent at high frequency.
Experimental data on bulk Nb in the GHz range show that $R_\mathrm{res,mag} \propto \sqrt{\omega} H_\mathrm{ext}$. $R_\mathrm{res,mag}$ calculated from Eq.~(\ref{eq:a35}) in the intermediate frequency range is given by 
\begin{equation}
R_\mathrm{res,mag} = \frac{H_\mathrm{ext}}{H_\mathrm{c}} \sqrt{\frac{\mu_0 \rho_\mathrm{n} \omega}{2g}},
\label{eq:a36}
\end{equation}
where $H_\mathrm{c}$ is the thermodynamic critical field and $g$ is a parameter related to the anisotropy of the superconductor. In the case of a 1.5~GHz Nb cavity ($\rho_\mathrm{n}\sim5\times10^{10}~\Omega\cdot\mathrm{m}^{-1}$, $H_\mathrm{c}=2000$~Oe, $2g=1$), the residual resistance due to the Earth's magnetic field ($\sim 0.5$~Oe) could be as high as $\sim 600$~n$\Omega$, about 30 times higher than $R_\mathrm{BCS}$ at 2.0~K. By applying magnetic shields around superconducting cavities, it is possible to shield external fields down to $\sim 1-10$~mOe.

\begin{figure}[ht]
\begin{center}
\includegraphics[width=8cm]{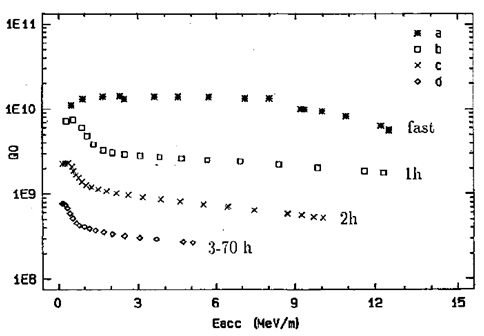}
\caption{The degradation of the quality factor in a 1.5~GHz bulk Nb cavity at 2.0~K after it was held in the region 100--150~K for various lengths of time~\cite{Bonin1991}.}
\label{fig:hydride}
\end{center}
\end{figure}

Another well-known contribution to the residual resistance in Nb cavities is due to the precipitation of normal-conducting niobium hydride islands near the surface, if the bulk H concentration is greater than $\sim 5 \times 10^{-4}$~wt.$\%$ and if the cool-down rate is $<1~\mathrm{K}\cdot\mathrm{min}^{-1}$ in the temperature range 75--150~K. Figure~\ref{fig:hydride} shows an example of $Q_0(E_\mathrm{acc})$ measured in a 1.5~GHz bulk Nb cavity at 2.0~K after it was held in the region 100--150~K for various lengths of time~\cite{Bonin1991}. This problem can be mitigated by degassing the cavity in a ultra-high-vacuum furnace at 600--800$^\circ$C for 2--6~h.
Tunnelling measurements of bulk Nb samples show the presence of electronic states within the energy gap. The density of states as a function of energy $N(\epsilon)$ obtained from the measurements can be described by the phenomenological model of Dynes~\cite{Dynes1978, Dynes1984}, resulting in a finite density of states at the Fermi level, $N(0)\simeq \gamma N_\mathrm{n}/\Delta$, where $\gamma$ is a damping parameter and $N_\mathrm{n}$ is the density of states in the normal state. The contribution to $R_\mathrm{res}$ from normal electrons occupying subgap states can be estimated from the two-fluid model surface resistance with $\gamma \sigma_n/\Delta$ replacing $\sigma_1$ \cite{Gurevich2012}:
\begin{equation}
R_\mathrm{res,subgap} \sim \mu_0^2 \omega^2 \lambda^3 \sigma_\mathrm{n} \gamma / \Delta,
\label{eq:a37}
\end{equation}
Residual resistance values of $\sim 10$~n$\Omega$, similar to those obtained in bulk Nb cavities, could result from Eq.~(\ref{eq:a37}) for $\gamma /\Delta \sim 10^{-3}$. The origin of these subgap states is not well understood. They could be due to intrinsic effects, such as strong impurity scattering or strong electron--phonon coupling, or extrinsic effects, such as normal-conducting precipitates, defective oxides, or pinned fluxoids.
\subsection{High-temperature superconductors}
\begin{figure}[ht]
\begin{center}
\includegraphics[width=8cm]{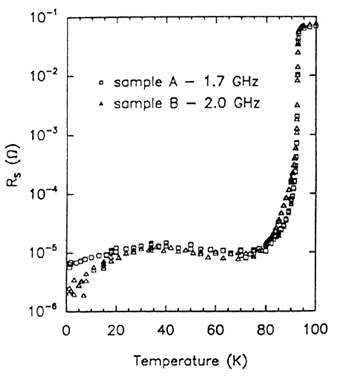}
\caption{$R_\mathrm{s}(T)$ measured on YBCO samples. Reprinted from~\cite{Hein1996} with permission from Nova Science Publishers, Inc.}
\label{fig:HTS}
\end{center}
\end{figure}

Superconductors with relatively low $T_\mathrm{c}$, such as Nb, have a `\textit{s}-wave' character, meaning that the energy gap is uniform in the momentum space. On the other hand, high-temperature cuprates ($T_\mathrm{c} \sim 90$~K) have a `\textit{d}-wave' character, indicating that the energy gap has four nodes along symmetric directions in momentum space. This leads to unfavorable consequences for RF application: the temperature dependence of the surface resistance follows a power law, instead of being exponential, and residual losses are higher. The $R_\mathrm{s}(H_p)$ dependence also exhibits strong non-linearity. The coherence length is much shorter than that of Nb (1--2~nm, instead of $\sim$~40 nm); therefore the pairing of electrons can easily be disrupted by defects. Cuprates are also `granular' superconductors, with high grain boundary resistance contributing to high $R_\mathrm{res}$. All these effects hinder their use for SRF cavity application where low $R_\mathrm{s}$ at high RF fields is required. Figure~\ref{fig:HTS} shows an example of $R_\mathrm{s}(T)$ measured in YBCO samples at 1.7 and 2~GHz~\cite{Hein1996}.
Superconductors with higher $T_\mathrm{c}$ than Nb but still \textit{s}-wave, such as Nb$_3$Sn ($T_\mathrm{c} \simeq 18$~K), NbN, and NbTiN ($T_\mathrm{c} \simeq 17$~K),  are more promising alternative materials for RF application. Their higher $T_\mathrm{c}$ value could allow operating cavities at 4.2~K instead of 2.0~K, therefore reducing the cost of refrigeration.

\section{The superheating field}
In the previous section, we discussed the physics determining the quality factor of SRF cavities at low RF fields. However, the feasibility of SRF cavities for particle accelerator applications, particularly those for high-energy physics, relies on the ability to reach high accelerating gradients, corresponding to high RF magnetic fields on the cavity surface. In this respect, it is important to understand what is the maximum RF magnetic field that can be applied to the surface of a superconductor (assumed to be `defect-free'), before a transition to the normal state occurs.

\begin{figure}[ht]
\begin{center}
\includegraphics[width=12cm]{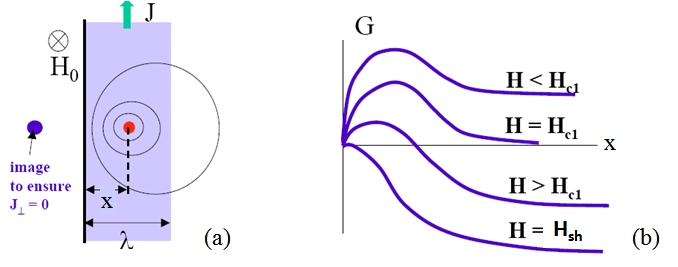}
\caption{(a) A schematic representation of a fluxoid near the surface of a superconductor. (b) The Gibbs free energy as a function of the fluxoid position for different applied fields.} 
\label{fig:vortex}
\end{center}
\end{figure}

Theoretically, the critical RF magnetic field is considered to be the so-called superheating field, $H_\mathrm{sh}$, which is the highest field up to which the superconductor remains in the Meissner state, and is unaltered by the dissipative motion of fluxoids. At $H_0=H_\mathrm{sh}$, the screening surface current reaches the depairing value, $J_\mathrm{d}=n_\mathrm{s} e \Delta/p_\mathrm{F}$, meaning that the kinetic energy of the superconducting carriers exceeds the binding energy of the Cooper pairs.
The magnetic field at which the Gibbs free energy has the same value whether the first fluxoid is inside or outside a type II superconductor is the so-called lower critical field, $H_\mathrm{c1}$. However, a fluxoid entering the surface of a superconductor has to overcome the so-called Bean--Livingston surface barrier~\cite{Bean1964}, which arises because of the attractive force between the fluxoid near the surface and its anti-fluxoid image, at the same distance from the surface but on the opposite side, required to ensure that the normal component of the current density at the boundary is zero. This is shown schematically in Fig.~\ref{fig:vortex}. The Gibbs free energy of a superconductor with a single fluxoid, as a function of the fluxoid position, is given by
\begin{equation}
G(x)=\phi_0 \left[ H_0 \mathrm{e}^{-x/\lambda} - \frac{\phi_0}{4\pi \lambda^2} K_0\left(\frac{2x}{\lambda} \right) + H_\mathrm{c1} - H_0 \right],
\label{eq:a38}
\end{equation}
where the first term is the energy due to the Meissner current, the second term is half of the interaction energy between the fluxoid and its image, the third term is the fluxoid self-energy, and the last term is the work done by the source of the applied field. $K_0$ is the zeroth-order Hankel function. The behaviour of $G(x)$ is shown schematically in Fig.~\ref{fig:vortex}: the surface barrier disappears only at $H_0 = H_\mathrm{sh} > H_\mathrm{c1}$.

Theoretical calculations of the superheating field as a function of the Ginzburg--Landau (GL) parameter, $\kappa_\mathrm{GL}$, close to $T_\mathrm{c}$ have been done since the 1960s using the GL theory~\cite{Ginzburg1958, deGennes1965, Matricon1967}. The metastability of the Meissner state leading to the existence of a superheating field represents a local minimum of the free energy, meaning that its second derivative is positive. The field up to which this metastability condition is satisfied has been evaluated considering fluctuations of the order parameter and the supercurrent along the boundary of the superconductor in one or two dimensions~\cite{Galaiko1966, Kramer1968}. $H_\mathrm{sh}$ resulting from the GL theory ($T \approx T_\mathrm{c}$) is given by
\begin{align}
\label{eq:a39}
\begin{split}
H_\mathrm{sh}\simeq 1.2 H_\mathrm{c}, \mbox{\hspace{10 mm}} \kappa_\mathrm{GL} \approx 1, \\
H_\mathrm{sh} = 0.745 H_\mathrm{c}, \mbox{\hspace{10 mm}} \kappa_\mathrm{GL} \gg 1.
\end{split}
\end{align}
The calculation of $H_\mathrm{sh}$ has recently been extended over the whole temperature range, $0<T<T_\mathrm{c}$, by the numerical solution of Eilenberger equations~\cite{Catelani2008, Pei-Jen2012}. These equations were obtained from Gorkov's formulation of the BCS theory. The calculations were done in the high-$\kappa_\mathrm{GL}$ limit and as a function of the mean free path. In the clean limit, $H_\mathrm{sh}$ at 0~K is given by
\begin{equation}
H_\mathrm{sh}=0.845 H_c, \mbox{\hspace{10 mm}} \kappa_\mathrm{GL} \gg 1.
\label{eq:a40}
\end{equation}
A weak dependence of $H_\mathrm{sh}$ on $\ell$ was found, if scattering was due to non-magnetic impurities, and an enhancement of $H_\mathrm{sh}$ (0~K) up to $\simeq 4.2\%$ at $T=0$ was found for $\pi \xi_0/\ell \simeq 0.6$~\cite{Pei-Jen2012}.

\begin{figure}[ht]
\begin{center}
\includegraphics[width=8cm]{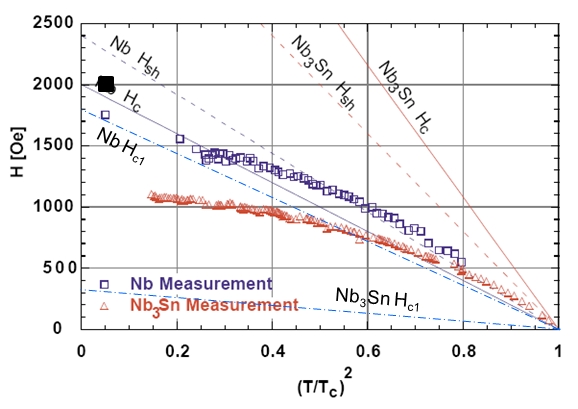}
\caption{Measurements of the maximum RF field achievable in bulk Nb and Nb$_3$Sn 1.3 GHz cavities as a function of temperature~\cite{Hays}. The solid, dashed, and dashed--dotted lines show the theoretical expectations for $H_\mathrm{c}(T), H_\mathrm{sh}(T)$, and $H_\mathrm{c1}(T)$, respectively. The solid black square shows the highest RF magnetic field ever measured on Nb cavities at 2.0~K.}
\label{fig:Hsh}
\end{center}
\end{figure}

Niobium, being a marginal type II superconductor ($\kappa_\mathrm{GL} \approx 1$), is a difficult material to study theoretically. Assuming the result in Eq.~(\ref{eq:a39}) to be valid at low $T/T_\mathrm{c}$, one can estimate the $H_\mathrm{sh}$ of Nb to be $\sim 240$~mT at 0~K. On the other hand, the $H_\mathrm{sh}$ of Nb$_3$Sn ($\kappa_\mathrm{GL} \approx 30$) obtained from Eq.~(\ref{eq:a40}) is of the order of 420~mT at 0~K.
Results from experiments aimed at measuring the $H_\mathrm{sh}$ of bulk Nb and of a thin Nb$_3$Sn layer grown on a Nb cavity at 1.3~GHz as a function of temperature are shown in Fig.~\ref{fig:Hsh}~\cite{Hays}. The data show that RF magnetic fields above $H_\mathrm{c1}$ have been achieved in both materials, but the highest field for Nb$_3$Sn is much lower than the value of $H_\mathrm{sh}$ predicted by the theory.

Unfortunately, the surface barrier can easily be suppressed by defects, such as nano-precipitates or even roughness, making it quite difficult in practice to extend the Meissner state up to $H_{sh}$ in high-$\kappa_\mathrm{GL}$ materials (usually deposited as thin films on a substrate).
In 2006, Gurevich proposed the use of multi-layered films of alternating superconductor--insulator--superconductor deposited on a bulk Nb cavity to achieve the superheating field on the cavity surface~\cite{Gurevich2006}. By making the thickness of the superconductor layer, \textit{d}, smaller than the penetration depth, the $H_\mathrm{c1}$ of the layer should increase significantly, as calculated by Abrikosov in 1964~\cite{Abrikosov1964}:
\begin{equation}
H_\mathrm{c1}=\frac{2 \phi_0}{\pi d^2} \left( \ln \frac{d}{\xi} - 0.07 \right).
\label{eq:a41}
\end{equation}

\section{$R_\mathrm{s}(H_0)$ dependence due to thermal feedback}
As we discussed in Sections 3 and 4, there are well-established theories to calculate the surface resistance and the RF critical field of superconductors. Unfortunately, there is no well-established theory describing the dependence of the surface resistance on the amplitude of the RF magnetic field, from low field up to $H_\mathrm{sh}$. In general terms, the surface resistance is expected to increase with increasing RF field because the density of thermally activated normal electrons increases. This occurs because the energy gap is reduced by the kinetic energy of the Cooper pairs to an effective gap $\Delta_\mathrm{eff}(v_\mathrm{s}) = \Delta - p_\mathrm{F} v_\mathrm{s}$, where $p_\mathrm{F}$ is the Fermi momentum and $v_\mathrm{s} = H_0/\lambda e n_\mathrm{s}$ is the superfluid velocity. A decrease of $Q_0$ with increasing $E_\mathrm{acc}$ is typically observed in SRF cavities, as shown, for example, in Fig.~\ref{fig:performance}. The measured slope of the $Q_0(E_\mathrm{acc})$ dependence changes significantly for different cavity treatments and for different cavity material (e.g. Nb bulk vs. thin film). In this section, we will discuss a thermal feedback model proposed by Gurevich~\cite{Gurev2006} as a simple mechanism causing an increase of $R_\mathrm{s}$ with increasing amplitude of the RF field.

\begin{figure}[ht]
\begin{center}
\includegraphics[width=12cm]{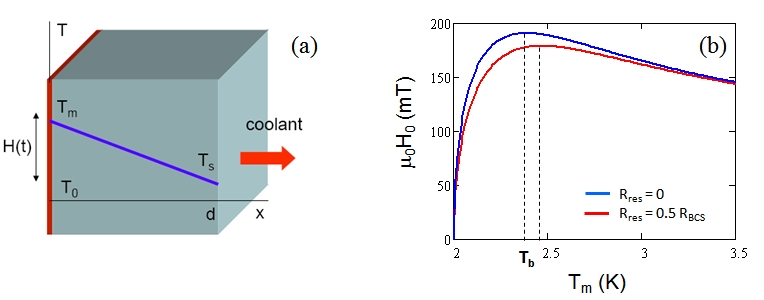}
\caption{(a) A schematic of heat transfer from the inner to the outer cavity surface. (b) The $H_0(T_\mathrm{m})$ dependence calculated from Eqs. (\ref{eq:a43}) and (\ref{eq:a44}) for Nb at 1.5~GHz and $T_0 = 2.0$~K. The maximum in the $H_0(T_\mathrm{m})$ defines the point of thermal breakdown.}
\label{fig:Tfeedback}
\end{center}
\end{figure}

Let us consider the one-dimensional case of a superconductor of thickness \textit{d} with a vacuum on one side (the inner surface of an SRF cavity) and liquid He on the other side (the outer surface of an SRF cavity). An RF field of amplitude $H_0$ is applied on the vacuum side, parallel to the surface. The RF power dissipated on the inner surface at temperature $T_\mathrm{m}$ is transferred as heat to the He bath of temperature $T_0$. The temperature of the outer surface is $T_\mathrm{s}$. The temperature profile is shown schematically in Fig.~\ref{fig:Tfeedback}. The heat balance equations for the inner and outer surfaces can be written as follows:
\begin{equation}
\frac{1}{2} R_\mathrm{s}(T_\mathrm{m})H_0^2 = \int_{T_\mathrm{m}}^{T_\mathrm{s}} \kappa(T) \,\mathrm{d}T = h_\mathrm{K}(T_\mathrm{s}, T_0)(T_\mathrm{s} - T_0),
\label{eq:a42}
\end{equation}
where $\kappa(T)$ is the thermal conductivity and $h_\mathrm{K}(T_\mathrm{s}, T_0)$ is the Kapitza conductance, which is the heat transfer coefficient between the outer surface and the superfluid He bath ($T_0 < 2.17~K$).

\begin{figure}[ht]
\begin{center}
\includegraphics[width=8cm]{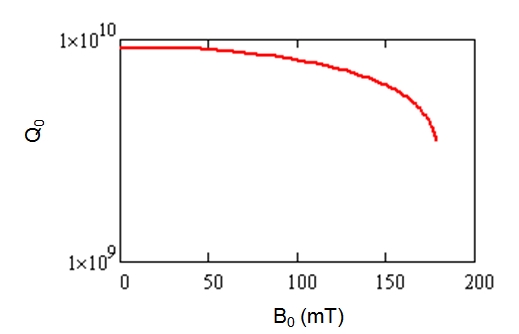}
\caption{$Q_0(B_0)$ calculated up to the thermal breakdown field for a 1.5~GHz Nb cavity at 2.0~K with the parameters mentioned in the text and $R_\mathrm{res}=10$~n$\Omega$.}
\label{fig:Q_curve}
\end{center}
\end{figure}

As will be shown below, the overheating of the inner surface is small relative to the He bath temperature ($T_\mathrm{m} - T_0 \ll T_0$); therefore we can neglect the temperature dependence of $\kappa$ and $h_\mathrm{K}$ and use their values at $T_0$. This simplifies Eq.~(\ref{eq:a42}) to the following:
\begin{equation}
\frac{1}{2} R_\mathrm{s}(T_\mathrm{m})H_0^2 = \frac{h_\mathrm{K} \kappa}{\kappa + d h_\mathrm{K}} (T_\mathrm{m} - T_0).
\label{eq:a43}
\end{equation}
By substituting the following approximate dependence of $R_\mathrm{BCS}(T_\mathrm{m})$,
\begin{equation}
R_\mathrm{BCS}(T_\mathrm{m}) \simeq \frac{A \omega^2}{T_\mathrm{m}} \mathrm{e}^{-\Delta/k_\mathrm{B} T_\mathrm{m}},
\label{eq:a44}
\end{equation}
in $R_\mathrm{s}(T_\mathrm{m}) = R_\mathrm{BCS}(T_\mathrm{m}) + R_\mathrm{res}$, Eq.~(\ref{eq:a43}) gives a relation between $H_0$ and $T_\mathrm{m}$. This dependence is plotted, as an example, in Fig.~\ref{fig:Tfeedback} for 3~mm thick Nb at 1.5~GHz, with $R_\mathrm{BCS}$(2~K) = 20~n$\Omega$, $\Delta /k_\mathrm{B} = 17.1$~K, $h_\mathrm{K} = 5~\mathrm{kW}\cdot\mathrm{m}^{-2}\cdot\mathrm{K}^{-1}$, $\kappa = 10~\mathrm{kW}\cdot\mathrm{m}^{-1}\cdot\mathrm{K}^{-1}$, and $T_0=2.0$~K. The maximum of the $H_0(T_\mathrm{m})$ curve corresponds to a thermal quench of the cavity. This point defines the temperature of the inner surface, $T_\mathrm{b}$, and the magnetic field, $H_\mathrm{b}$, at which thermal breakdown occurs. $T_\mathrm{b}$ is the value of $T_\mathrm{m}$ that satisfies $\mathrm{d}H_0/\mathrm{d}T_m = 0$, where $H_0(T_\mathrm{m})$ is given by Eq.~(\ref{eq:a43}), with Eq.~(\ref{eq:a44}) for $R_\mathrm{BCS}$. Neglecting the residual resistance, one obtains
\begin{equation}
T_\mathrm{b} - T_0 \approx \frac{T_0^2}{\Delta}.
\label{eq:a45}
\end{equation}
where $T_\mathrm{b} - T_0 \approx 0.23$~K for Nb at 2.0~K.
Substituting Eq~(\ref{eq:a45}) into Eq.~(\ref{eq:a43}), one obtains
\begin{equation}
H_\mathrm{b}^2 \approx \frac{2 k_\mathrm{B} T_0^2 h_\mathrm{K} \kappa}{(\kappa + d h_\mathrm{K}) \Delta e R_\mathrm{BCS}},
\label{eq:a46}
\end{equation}
where $e=2.718$.
The breakdown field estimated from Eq.~(\ref{eq:a46}) is that which occurs in the case of uniform heating of the inner surface, without any localized defect. Because both $H_0$ and $R_\mathrm{s}$ depend on the inner surface temperature, Eqs.~(\ref{eq:a43}) and (\ref{eq:a44}) provide a dependence of $R_\mathrm{s}$ on $H_0$, and it is therefore possible to calculate a $Q_0(H_0)$ curve ($Q_0=G/R_\mathrm{s}$). This is shown, as an example, in Fig.~\ref{fig:Q_curve} for a 1.5~GHz bulk Nb cavity with $G=280$~n$\Omega$ at 2.0~K and the thermal parameters mentioned earlier.

For a thin film of thickness $d_\mathrm{tf}$ and thermal conductivity $\kappa_\mathrm{tf}$ deposited on a substrate, the total thermal conductance becomes $h_\mathrm{K}/[1+h_\mathrm{K}(d/\kappa + d_\mathrm{tf}/\kappa_\mathrm{tf})]$. In the case of a 1.5~$\mu$m thick Nb$_3$Sn film [$\kappa_\mathrm{tf} \simeq 10^{-2}~\mathrm{W}\cdot\mathrm{m}^{-1}\cdot\mathrm{K}^{-1}$] on top of a 3~mm thick Nb cavity, the heat transfer coefficient of the thin film, $d_\mathrm{tf}/\kappa_\mathrm{tf} \simeq 670~ \mathrm{W}\cdot\mathrm{m}^{-2}\cdot\mathrm{K}^{-1}$, is comparable to that of the Nb substrate, $d/\kappa \simeq 330~\mathrm{W}\cdot\mathrm{m}^{-2}\cdot\mathrm{K}^{-1}$, at 2.0~K.

\section{Summary}
Unlike the DC case, superconductors in an RF field have a non-zero surface resistance.
The surface resistance can be easily understood in terms of a two-fluid model and is due to the interaction of the electric field (exponentially decaying from the surface) with normal-conducting electrons.

At low RF fields, the surface resistance can be expressed as the sum of the BCS surface resistance and a residual resistance. The low-temperature BCS surface resistance:
\begin{itemize}
\item increases quadratically with frequency,
\item decreases exponentially with temperature, and
\item has a minimum as a function of the material purity.
\end{itemize}
There are many possible contributions to the residual resistance, and trapped fluxoids and niobium hydride precipitates have been proven to be among them.

The maximum theoretical RF field that can be applied to the surface of a superconductor is the superheating field, which is of the order of the thermodynamic critical field. Superconductor--insulator--superconductor multilayer thin films might be a possible way to reach the superheating field for superconducting materials other than Nb.

There exists no well-accepted theory of the surface resistance at high RF fields that could be used to describe the $Q_0(E_\mathrm{acc})$ curves of SRF cavities. Thermal feedback between RF power dissipation and surface temperature is a simple extrinsic mechanism causing an increase of the surface resistance with increasing RF field.

\section*{Acknowledgements}

I wish to thank Professor A. Gurevich of Old Dominion University for many fruitful discussions and for providing some of the figures.

\section*{Bibliography}

H. Padamsee. \emph{RF Superconductivity: Science, Technology, and Applications} (Wiley-VCH, Weinheim, 2009).
H. Padamsee, J. Knobloch, and T. Hays. \emph{RF Superconductivity for Accelerators}
(John Wiley \& Sons, New York, 1998).

\end{document}